# Uncertainty Quantification and Composition Optimization for Alloy Additive Manufacturing Through A CALPHAD-based ICME Framework

**Xin Wang, Wei Xiong** [*]

*Physical Metallurgy and Materials Design Laboratory,
Department of Mechanical Engineering and Materials Science,
University of Pittsburgh, Pittsburgh, PA 15261, USA*
[*] *Corresponding Author, Email: weixiong@pitt.edu, w-xiong@outlook.com*

## Abstract

During powder production, the pre-alloyed powder composition often deviates from the target composition leading to undesirable properties of additive manufacturing (AM) components. Therefore, we developed a method to perform high-throughput calculation and uncertainty quantification by using a CALPHAD-based ICME framework (CALPHAD: calculations of phase diagrams, ICME: integrated computational materials engineering) to optimize the composition, and took the high-strength low-alloy steel (HSLA) as a case study. We analyzed the process-structure-property relationships for 450,000 compositions around the nominal composition of HSLA-115. Properties that are critical for the performance, such as yield strength, impact transition temperature, and weldability, were evaluated to optimize the composition. With the same uncertainty as the initial composition, an optimized average composition has been determined, which increased the probability of achieving successful AM builds by 44.7%. The present strategy is general and can be applied to other alloy composition optimization to expand the choices of alloy for additive manufacturing. Such a method also calls for high-quality CALPHAD databases and predictive ICME models.

# Introduction

The ability to produce complex geometries, the capability of processing small batches with low cost, and the capacity to perform *in-situ* repair, makes alloy additive manufacturing (AM) a market worth billions of dollars[1]. In alloy AM, the feedstock is melted by a heat source such as a laser or electron beam to build the parts layer by layer[2]. Parameters like part geometry, scan strategy, build chamber atmosphere, and feedstock properties, are factors that directly impact the performance of the AM components[3]. The inherent uncertainties in these parameters lead to an unavoidable variation in quality[4]. As a result, the confidence in the quality of AM products is low due to a lack of uncertainty quantification and design sensitivity analysis, which is impeding the commercialization of alloy AM[5].

Uncertainty quantification is essential for quality control in manufacturing. Based on the given uncertainty of processing parameters, uncertainty quantification can determine the variation in microstructure, and mechanical properties for AM builds. Currently, most of the uncertainty quantification studies focus on manufacturing processes[4,6,7]. However, the influence of uncertainty in the chemical composition of feedstock is often overlooked. The cost of metal materials is the second-highest in AM part[8], and the feedstock quality plays a vital role in the AM builds performance. Deviation from the desired composition could lead to detrimental phase formation during solidification[9] and post-heat treatment[10]. It may also introduce cracks, pores, and alter physical properties such as specific heat and melting point, which will further influence the choice of processing parameters[11]. Moreover, the composition variation in AM products is unavoidable, which comes from various sources. And three common causes can be identified. First, there is a variation in the composition of the powders manufactured in different batches[12]. Second, the composition of the AM build will deviate locally from the nominal composition of the feedstock[13]. Third, to improve sustainability, a large amount of unprinted powder needs to be recycled after the AM process, which causes degradation with contamination. The above composition related issues propagate uncertainty throughout the AM process and should be addressed during the composition design of the feedstock material. This implies that the nominal composition of an alloy needs to be well designed to avoid the negative impact of the uncertainty on the final build. However, the study of the correlation between alloy composition and performance of AM builds after post-treatment is limited. Only a few studies have reported the impact of composition variation on the AM builds with experiments[14,15]. Moreover, the comprehensive modeling tool to facilitate decision making on the composition range in feedstock manufacturing is yet unavailable.

The ICME (Integrated Computational Materials Engineering) method can solve this problem by determining the allowable variation from the intended composition based on the process-structure-property relationships[16,17]. The ICME method will reduce the dependence on experimental trials, and thus accelerate the materials design[11]. In order to address the issue of the composition uncertainty of feedstock, it is critical to establish an ICME model framework to simulate the process-structure-property relationships in alloys. In this work, high-strength low-alloy (HSLA)-115 (115 corresponds to minimum achievable tensile yield strength in ksi, which is equivalent to 793 MPa) steel was chosen to demonstrate the effectiveness of this design framework. We expected that with the implementation of uncertainty quantification through such an ICME model framework, the nominal composition of the cast HSLA-115 steel could be optimized to increase the likelihood of successful AM builds, which should meet all the property requirements after heat treatment. Initially, we determined the process-structure-property relationship of HSLA steel to model the properties as a function of composition. Further, the following models were applied for predicting the properties: 1) CALPHAD (Calculation of Phase Diagrams) method[18] in combination with phenomenological models for predicting the dislocation density[19], grain size[20,21], impact transition temperature (ITT)[22], and carbon equivalent[23]; 2) Data-mining decision tree model for martensite start ($M_S$) temperature[24]; and 3) Physics-based strengthening model[17] consisting of the simulation of hardening effect caused by dislocations[25], grain boundaries[26,27], precipitates[28,29] and solid solution atoms[30,31] to predict the yield strength, low-temperature ductility, and weldability for a given composition and heat treatment process. Finally, high-throughput calculations were performed for a range of compositions to optimize the nominal composition of cast HSLA-115 steel for AM. By employing the ICME framework developed in this work to optimize the composition of the HSLA-115 steel powders, the probability of achieving the desired properties in the AM build with heat treatment increases significantly.

# Results and discussion

**HSLA-115 steel for AM**

The HSLA steels are widely used in many structural applications, such as bridges, ship hulls, and mining equipment[32–35]. Due to the excellent mechanical properties and good weldability, HSLA steel is an outstanding candidate for AM. Unlike Inconel 718, Ti-64, and Stainless Steel 316, there is no commercial AM powder available. The HSLA alloy powder production is in a small batch and very expensive. Thus, it is necessary to develop a method to optimize the alloy composition to make sure the powder we customized can be printed without cracks and have the desired property. The composition and its uncertainty for typical wrought HSLA-115 steel are listed in Table 1. The composition uncertainty range was specified by the powder vendor, Praxair, Inc., for the HSLA-115 steel powder manufactured for laser powder AM.

The process-structure-property relationships for high-performance AM HSLA steels are summarized in the systems design chart, as shown in Fig. 1. The systems design chart exhibits how hierarchical structural features contribute to the mechanical properties and how the structure evolves during different processes and compositions[17,36]. Each line connecting the process, structure, and property indicates a relationship/model between these attributes. HSLA steel has a combination of high strength and good low-temperature impact toughness. This is achieved through hot isostatic pressing (HIP)/austenitization, quenching, and tempering that leads to a dense part with a fine martensite/bainite matrix and dispersed nano-sized Cu and $M_2C$ precipitates. HIP aims to reduce the porosity of as-built components for improved mechanical properties as well as corrosion resistance[37]. In dense builds, austenitization helps in achieving homogenized austenitic structure with the dissolution of undesirable phases and elimination of segregation due to rapid solidification. During post-heat treatment, it is expected that enough undissolved MX particles (mainly the NbC) exist to pin the grain boundaries and prevent excessive grain growth. Water quenching is applied to form a fine lath bainitic/martensitic structure that improves the strength. Lastly, the tempered martensite formed after tempering enhances the impact toughness with the reduction in dislocation density. More importantly, the coherent Cu (3-5 nm in radius) and $M_2C$ (1.5-3 nm in radius) will precipitate during tempering, causing the major hardening effect[38,39]. The precipitation of $M_2C$ will dissolve the cementite and avoid the decrease in impact toughness due to the formation of coarse cementite. Other precipitates such as $M_{23}C_6$ may also form while they usually have large size and contribute negligible strengthening effect[40]. Finally, the good weldability of this steel originates from the low content of carbon and other alloying elements[41].

**Process-structure-property models used in the ICME framework**

In this work, an ICME framework has been established to evaluate the yield strength, weldability, and impact transition temperature of HSLA steels based on the systems design chart shown in Fig. 1. As illustrated in Fig. 2, the composition and processing parameters were taken as inputs for the decision tree model, CALPHAD-based thermodynamic model, and Graville diagram[23]. The outputs from these models, such as the dislocation density, matrix composition, and etc. were coupled with the physics-based strengthening, ITT, and weldability evaluation models to calculate the yield strength, ITT, and weldability that includes the freezing range and Graville diagram index for each composition. Finally, the calculated properties for each composition were used to find the optimized composition for AM that will give the highest chance of a successful build that meets all property requirements. The explicit description of models, the screening and analysis process can be found in the method section.

**Composition screening analysis**

Figure 3 shows the model predicted yield strength against the experimental measurements for several HSLA steels[38,42,43] with different compositions and tempering temperature ranges from 450 to 650 °C (For alloys heat treated with the same temperature and different time, the closest value to prediction was chosen in Fig. 3). The ICME model prediction and experimental results show a good agreement. These results indicate that the strengthening model within the ICME framework is capable of predicting the yield strength of HSLA steels.

Figure 4 shows the variation of all properties as a function of carbon content. The same procedure is also applied to other elements. It allows us to assess the influence of each element on the strength, low-temperature ductility, and weldability. Each column represents the model prediction for one set of compositions with the same range of carbon content, i.e., 0.0025 wt.% carbon. The number under each bin corresponds to the smallest carbon content in the bin. For example, bin 0.05 contains all compositions that have the carbon content between 0.05 and 0.0525, i.e., [0.050, 0.0525) and other elements in their initial composition range, which is listed in Table 1. Evidently, with the increase in carbon content, the yield strength, as shown in Fig. 4(a), initially increases and then decreases, which is different from Saha's[44] work on the high-strength steels that the strength will continuously increase with the addition of carbon content. The contradiction is from the incorrect assumption in Saha's work that carbon only forms the $M_2C$. However, carbon will also dissolve in the martensite matrix, and form other carbides. Furthermore, the fraction of $M_2C$ will change with different compositions. Based on our calculation (Supplementary Figure 1), when the carbon is increasing, the fraction of $M_2C$ will increase first and then decrease. Moreover, even if the carbon content is similar, the fraction of $M_2C$ will also change with different alloying elements. For low-temperature ductility, as the carbon content increases, the ITT increases and then decreases (see Fig. 4(b)), indicating worsening of low-temperature ductility at the first stage and improvement in the later stage. Further, as more carbon is added to an alloy, the freezing range increases, as shown in Fig. 4(c), which indicates a higher probability of hot cracking. Similarly, the location of the composition in the Graville diagram will move out of Zone I when the carbon content is around 0.085 wt.%, and the susceptibility to cold cracks increases, as shown in Fig. 4(d). These results are consistent with the expected influence of carbon content on the weldability of HSLA steels.

The influence of carbon content on the yield strength and different hardening effects are shown in Fig. 5(a). The increase in carbon content leads to an increase in strengthening effects from grain boundaries and dislocations since carbon introduces the formation of Zener pinning particle NbC, and promotes higher dislocation density after quenching. However, the strength achieved from the precipitation hardening increases initially and then decreases, which results in a peak hardening with the carbon content between 0.06-0.065 wt.%. Precipitation hardening is critical, and it depends on the formation of nano-size $M_2C$ and Cu particle in the HSLA-115 steel. According to Fig. 5(b), the addition of carbon has no apparent influence on Cu precipitation, while it has a significant impact on the precipitation of $M_2C$.

Figure 6 provides an overview of the qualified composition range with all the considered properties as the selection criterion. In such a histogram, the composition sets of every single bin are categorized into different groups based on the number and type of criterion the composition meets. The percentage of compositions in the group with no pattern and in pink (compositions meeting all property requirements) continues to increase with the increase in carbon content, displaying a maximum at 0.06 wt.% carbon, which is higher than the initial nominal composition 0.053 wt.% carbon that is determined based on the cast HSLA steel. However, when the carbon content is higher than 0.085 wt.%, only a few compositions can satisfy the weldability requirement. Since the uncertainty in carbon content is ± 0.025 wt.%, it is better to avoid the targeted average carbon content higher than 0.0575 wt.%. The insufficient strength in this composition range can be made up by tuning the composition of other elements to increase the hardening effects.

Other elements were screened and analyzed using the same method that was implemented for carbon. In total, 450,000 compositions sampled using the strategy mentioned in the method section were calculated and analyzed. Table 2 summarizes the elemental influence on structure and strengthening effects within the composition range listed in Table 1. For instance, when molybdenum increases from 0.2 to 1.2 wt.%, the weldability continue to decrease. While the yield strength increases at first due to the improvement in precipitation hardening from $M_2C$ particles, solid solution strengthening, and dislocation hardening effects, and then decreases due to the reduction in the phase fraction of $M_2C$ when Mo reaches to a threshold value. The low-temperature ductility will firstly decrease and then increase. The influences of other elements can be explained based on Table 2 and following the same method.

Table 3 lists the initial and optimized composition in wt.%. In comparison with the initial composition, the contents of C, Cu, and Mo have increased to ensure that the yield strength is higher than 115 ksi (793 MPa), while the contents of Cr, Mn, and Si have decreased to balance the deterioration of weldability. The Nb content is increased to introduce a higher phase fraction of MX during the austenitization process to effectively avoid excessive grain growth, improve the low-temperature ductility, and increase the strength. Elements such as Mo, Ni, and Al do not change since their initial content is sufficient for the required properties, or they do not have a pronounced influence on critical properties.

**Verification of composition optimization**

In comparison with the calculated properties of the initial and optimized nominal composition (Table 4), it is evident that the optimized one has much higher yield strength and lower ITT from model prediction. This indicates that by slightly tuning the initial composition, HSLA steel could achieve a higher strength while remaining ductile at low temperatures. For example, less $M_{23}C_6$ and more $M_2C$ precipitates form at the tempering temperature with the optimized composition, as shown in Fig. 7. Also, a higher fraction of NbC remains stable at the high temperature and hence, retard the grain growth and coarsening. Importantly, the optimized alloy has achieved a small freezing range, and it is located in Zone I of the Graville diagram. This indicates that the printability for AM of alloy with initial composition is similar to the one after composition optimization.

To further verify the improvement after optimization in terms of the composition uncertainty, 50,000 compositions were randomly sampled from the initial and optimized composition spaces listed in Table 3, respectively. The yield strength, ITT, freezing range, and Graville diagram location were calculated for each data point. The same criteria listed in the previous sections were used to evaluate whether the composition meets the property requirements. According to Fig. 8, the optimized composition exhibits higher strength and lower ITT without sacrificing the weldability. Most importantly, the lowest strength and highest ITT among the 50,000 samples taken from the optimized composition with uncertainty are still higher than 115 ksi (793 MPa) and lower than 0°C, respectively. As a result, the optimized composition shows a higher chance of achieving successful builds (99.996%) compared with the initial composition (55.266%). Figure 9 illustrates how the composition was shifted to gain the highest success rate with a fixed composition uncertainty. In the composition space, there is a subspace that can meet all the required properties. However, all the initial composition with variation may not be present in that subspace, i.e., with the deviation from nominal composition, the AM build may not have the required properties. After the optimization, the nominal composition is shifted, and as a result, all the possible compositions meet the requirements taken into account with composition uncertainty.

In summary, our goal of maximizing the chance of a product meets all performance requirements in the presence of uncertainty is achieved. It is noted that this work is a reliability-based design optimization[45], while the robust design optimization that makes the product performance insensitive to the input uncertainties is beyond the scope of this study[46]. However, the mean and standard deviation of optimal composition determined using surrogate models are available in the supplementary files, which can be used to perform the robust design optimization.

**Limitation of current study and discussion on the uncertainties sources**

The uncertainty involved in this work includes the aleatory and epistemic uncertainty[6]. The aleatory uncertainty refers to natural variation and is hard to be avoided, while the epistemic uncertainty is originated from the lack of knowledge and approximations made in the modeling method[47]. In this work, our primary goal is to study the influence of the aleatory uncertainty of composition in the performance of post-heat treated AM builds and optimize the composition to gain higher chances of a successful build. We believe that based on the widely accepted physical models and the reliable databases developed several decades, such as the TCFE steel database released by the Thermo-Calc software company[48,49], the ICME model-prediction is effective to guide the composition optimization. However, it is noteworthy that the accuracy of ICME model-prediction relies on the quality of the CALPHAD database. Therefore, instead of performing a composition design based on the model-prediction with absolute values, we would rather aim at composition optimization by predicting the alloying effects with the trend analysis.

Due to the lack of experimental studies of the influence of composition change on AM build property, the uncertainty quantification for other uncertainties sources is challenging and is not performed in this work[50]. Other aleatory uncertainties from the processing parameters play an important role in the performance of AM. It should be further studied by coupling the CALPHAD-based ICME framework with the existing AM simulation models to address the process uncertainties in the future[4,51,52].

Epistemic uncertainty includes data uncertainty and model uncertainty[53]. Data uncertainty is originated from the limited information of data and is reduceable by collecting more data. In this study, only the composition uncertainty range is available, while the composition distribution is yet unknown. As a result, we have assumed that the distribution of composition is uniform. Since provided the composition within the uncertainty range meets the property requirement, the manufactured component should meet the property requirement regardless of the distribution. However, the calculated success rate may vary with different distributions. In the future, if massive production and chemical analysis are performed to gain the composition distribution, a more representative result can be obtained.

The other epistemic uncertainties represent the difference between model prediction and experimental observation is called model uncertainty, which includes the model form uncertainty, solution approximation, and model parameter uncertainty[4]. Model form uncertainty stems from the assumption/simplification in the model. For example, it is assumed that the precipitate size is a fixed value in this work, while the real precipitate size is within a range and follows a specific distribution. It is possible to further increase the model accuracy by simulating precipitate size distribution[54] and incorporating the size distribution into the strengthening model[55]. Moreover, the porosity also determines the strength and ductility, while the influence of porosity is not considered in the current model framework. The model bias can be further reduced by integrating the porosity prediction into the ICME model framework[56].

The model parameter uncertainty is originated from the fact that some parameters used in the model are not accurate enough. For example, the Hall-Petch coefficient used in this work is determined from references[57], while it may not be precisely the same for the alloy composition studied in this work, and it may lead to a discrepancy between the model output and experiments. Such uncertainties can be minimized by performing experiments to measure the parameters or collecting more literature to gain a more robust understanding of the value of these parameters. The last one is solution approximation. As shown in the supplementary files (Table S2-3), there are discrepancies between the surrogate models and original models. Such uncertainties can be reduced by generating more and unbiased training data, optimizing the hyper-parameters, etc. Finally, it is possible to calibrate models by experiment design and use the framework proposed by Kennedy and O'Hagan[58–60]. During the calibration process, the difference between model prediction and experiments can be modeled as a Gaussian process model, and the unknown model parameters shall be studied using the Bayesian calibration method[60].

In summary, the present study establishes an ICME framework for the composition optimization of AM HSLA steel powder. The major conclusions are as follows: An ICME framework supported by the CALPHAD model-prediction, phenomenological models, and physics-based models has been established for predicting the weldability, yield strength, and low-temperature ductility of AM HSLA with a given composition and post-treatment process without many structural information from experiments. In total, 450,000 compositions have been calculated using the ICME framework to identify the optimum composition, while taking into account the uncertainty, that can maximize the rate of a successful AM build. The proportion of alloys that meet the requirements for weldability, yield strength, and low-temperature ductility increased from 55.266% to 99.996% after optimization. This method can help transform the commercial alloys manufactured by conventional methods to the AM process. This optimization process is established for a general-purpose of composition optimization for additive manufacturing, but it can be further customized for a specific manufacturing process by integrating more process-structure-property models due to different types of processing. Although some of the ICME models adopted in this framework may be further improved depending on the alloy systems, the optimization strategy and concepts can be readily extended to other case studies. The composition range generated using this ICME framework is expected to be more reliable than the arbitrary range proposed by the powder vendor based on their experience.

## Methods

### ICME model framework

All the models were implemented using the TC-Python toolkit from Thermo-Calc software[61,62] and Python[63]. The following section will introduce the model framework in detail.

As illustrated in [Fig. 2](), the alloy yield strength, $\sigma_Y$[17], arises from the combined strengthening effects of Peierls-Nabarro (P-N) stress $\sigma_0$, dislocation strengthening $\sigma_d$, solid solution strengthening $\sigma_{ss}$, precipitation strengthening $\sigma_{ppt}$, and grain boundary strengthening $\sigma_{H\text{-}P}$:

$$\sigma_Y = \sigma_0 + \sigma_d + \sigma_{H-P} + \sigma_{ppt} + \sigma_{ss} \tag{1}$$

where $\sigma_0 = 50$ MPa is the P-N stress of α-Fe[21], the details of calculation for other strengthening effects are given as below.

The martensitic/bainitic structure in the HSLA-115 steel, with high dislocation density, forms due to rapid cooling. Takahashi and Bhadeshia[19] proposed a phenomenological equation to describe the relationship between the $M_S$ temperature, dislocation density, and the strengthening effect from dislocations in the as-quenched steel $\sigma_{DS}^0$)[19,25]:

$$\sigma_{DS}^0 = M\tau_{DS}^0 = 0.38MGb\sqrt{\rho} \tag{2}$$

$$\log(\rho) = 9.2848 + 6880.73/T - 1780360/T^2 \tag{3}$$

where temperature $T$ is max(570 K, $M_S$), and $M$ is the Taylor orientation factor to convert the shear stress to normal stress which ranges from 2.6 to 3.06 in bcc materials, and $M$ is 2.75 in this study[64], $G$ = 80 GPa is the shear modulus[65], $b$ = 0.25 nm is the Burgers vector in α-Fe[66], $\rho$ is the dislocation density. The $M_S$ temperature can be either predicted using theoretical modeling or determined using experiments such as dilatometry. In this work, we apply a data-mining generated decision tree model[24] for the prediction of $M_S$ temperature. The dislocation density will decrease during the tempering heat treatment, and it is related to the ratio of the precipitate fraction formed during the heat treatment process to the equilibrium value $f_{ppt}$[67]:

$$\sigma_{DS} = M\tau_{DS} = M(\tau_{DS}^0 - \sqrt{0.8}\tau_{DS}^0 f_{ppt}) \tag{4}$$

Since the fraction of precipitates does not increase significantly after a certain aging time[38,68], it is assumed that the ratio $f_{ppt}$ is 1 after tempering.

The contribution from solid solution strengthening arises from the size and elastic modulus misfit between the solvent and the solute atoms. Fleischer's equation[30] is adopted to evaluate the strengthening effects in multicomponent solid solutions[69]:

$$\sigma_{ss} = \left[\sum_i \beta_{ss,i}^2 c_i\right]^{0.5}, i = \text{Ni, Mn, Cr, Al, Mo, Cu} \tag{5}$$

where $k_{ss,Ni}$=708 MPa atomic fraction$^{-1}$ (MPa at$^{-1}$), $k_{ss,Mn}$=540 MPa at$^{-1}$ $k_{ss,Cr}$=622 MPa at$^{-1}$, $k_{ss,Al}$=196 MPa at$^{-1}$, $k_{ss,Mo}$=2362 MPa at$^{-1}$, $k_{ss,Cu}$=320 MPa at$^{-1}$ are the strengthening coefficients[31], and $c_i$ is the atomic fraction of the strengthening element in the matrix at the tempering temperature obtained using the CALPHAD method.

The most critical strengthening mechanism in HSLA-115 steel is the precipitation hardening due to Cu and M$_2$C precipitates at the tempering temperature (550°C). For predicting the strengthening effect of Cu precipitates, the Russel-Brown model is valid[28,70]. This model is based on the interaction between the dislocations and Cu precipitates, which originates from the difference in elastic modulus between the matrix and precipitates[28]:

$$\sigma_{Cu} = 0.8M \frac{Gb}{L_{Cu}} [1 - (\frac{E_p}{E_m})^2]^{\frac{1}{2}}; \quad \sin^{-1}\left(\frac{E_p}{E_m}\right) \leq 50° \tag{6}$$

$$\sigma_{Cu} = M \frac{Gb}{L_{Cu}} [1 - (\frac{E_p}{E_m})^2]^{\frac{3}{4}}; \quad \sin^{-1}\left(\frac{E_p}{E_m}\right) \geq 50° \tag{7}$$

where $E_p$ and $E_m$ are the dislocation line energy in the Cu precipitates and the matrix, respectively. $L_{Cu}$ is the mean planar spacing of Cu precipitates, and $L_{Cu}^{-1} = f_{Cu}^{\frac{1}{2}}/1.77 r_{Cu}$, $f_{Cu}$ is the volume fraction of Cu precipitates, and $r_{Cu}$ is the mean radius of the Cu precipitates. The $f_{Cu}$ is calculated using the Thermo-Calc software with the TCFE9 database. In order to simplify the model, we assume that the $r_{Cu}$ is 4 nm since the radius of the Cu-rich precipitate in aged HSLA alloy is usually 3-5 nm[40,43,68]. The $E_p/E_m$ ratio can be calculated with the following equations:

$$\frac{E_p}{E_m} = \frac{E_P^\infty \log \frac{r}{r_0}}{E_m^\infty \log \frac{R}{r_0}} + \frac{\log \frac{R}{r}}{\log \frac{R}{r_0}} \tag{8}$$

where $E_p^\infty$ and $E_m^\infty$ denote the energy per unit length of dislocation in an infinite medium, and their ratio is 0.62, $R = 1000 r_0$ is the outer cut-off radius, $r_0 = 2.5b$ is the inner cut-off radius or dislocation core radius[70].

The strengthening mechanism of M$_2$C precipitates in HSLA steels or similar alloys should follow the Orowan-Ashby dislocation strengthening effect, provided the precipitate size is larger than 1.1 nm[29]. For HSLA steel aged at 550°C, the mean radius of M$_2$C precipitate $r_{M2C}$ is usually less than 2.5 nm[38,68], and it is assumed that $r_{M2C} = 2$ nm in this work. The Orowan equation can be written in the following format[71]:

$$\sigma_{M2C} = MY \frac{G}{4\pi(1-v)^{0.5}} \frac{2b}{\omega_L r_{M2C}} \ln\left(\frac{2\omega_D r_{M2C}}{b}\right) \sqrt{\frac{\ln(\frac{2\omega_D r_{M2C}}{b})}{\ln(\frac{\omega_L r_{M2C}}{b})}} \tag{9}$$

$$\omega_L = \left(\frac{\pi \omega_q}{f_{M2C}}\right)^{0.5} - 2\omega_r \tag{10}$$

$$\frac{1}{\omega_D} = \frac{1}{\omega_L} + \frac{1}{2\omega_r} \tag{11}$$

where $v = 0.3$ is the Poisson's ratio, $Y = 0.85$ is the M₂C spatial-distribution parameter for Orowan dislocation looping, $f_{M2C}$ is the volume fraction of M₂C, $\omega_r$ is the constant to convert the mean particle radius of M₂C to the effective radius that intersects with the glide plane, and $\omega_q$ establishes the relationship between the mean area of precipitate intersecting with the glide plane. A detailed discussion about $\omega_r$, $\omega_q$, and $Y$ can be found in Ref.[38].

The following equation is used to evaluate the overall strengthening due to precipitation with two different sets of precipitates:

$$\sigma_{ppt} = (\sigma_{Cu}^k + \sigma_{M2C}^k)^{\frac{1}{k}} \tag{12}$$

where $k = 1.71$ is the superposition exponent to superpose the strengthening effects of two different strengthening particles[29].

The strengthening effect due to the grain size refinement can be estimated using the Hall-Petch equation[26,27]:

$$\sigma_{H-P} = \frac{k_y}{\sqrt{d_{packet}}} \tag{13}$$

where $k_y = 600$ MPa μm$^{-0.5}$ is the Hall-Petch coefficient[57], $d_{packet}$ is the size of the martensite packet or bainite which is closely related to the size of prior austenite $D_g$[21]. In a lower bainite/martensite matrix materials, the martensite block size will be even smaller[72,73]. As a result, we assume the grain size relationship is similar in martensitic steel, which can be written in the form of the following equation[21]:

$$d_{packet} = 0.40 D_g \tag{14}$$

The NbC phase in HSLA steels remains undissolved at the austenitization temperature (950°C), which can pin the austenite grain boundary to prevent excessive grain growth. The maximum austenite grain size after austenitization is a function of the size and volume fraction of pinning particles[20]:

$$D_g = \begin{cases} 8r_{MX}/(9f_{MX}^{0.93}), & f_{MX} < 0.1 \\ 3.6r_{MX}/(f_{MX}^{0.33}), & f_{MX} > 0.1 \end{cases} \tag{15}$$

where $r_{MX}$ is the average radius of the MX (M = Nb, X = C, N) in HSLA steels, and it is reported to be around 13 nm in different HSLA steels with various compositions and heat treatment parameters[74,75], $f_{MX}$ is the volume fraction of MX at austenitization temperature (e.g., 950°C) which can be obtained using the Thermo-Calc software with TCFE9 database.

As shown in Fig. 2, the ITT is used as an evaluation criterion for the low-temperature ductility. The ITT corresponds to the ductile-brittle transition temperature (DBTT) or fracture appearance transition temperature (FATT), which are close to each other. At a temperature above the ITT, the material is ductile; otherwise, it is brittle. The phenomenological equation to calculate 50% ITT[22] (°C) for the ferritic-pearlitic steels after the calibration with reported HSLA ITT[76] is given below:

$$50\% \text{ FATT} = 112t^{0.5} - 13.7d^{-0.5} + 0.43\Delta y - 54 \qquad (16)$$

where $t$ is the cementite thickness in μm, $d$ is the grain size in mm, $\Delta y$ is the strength contributed from the precipitation hardening in MPa that can be obtained through the precipitation strengthening model and Zener pinning effect as shown in Fig. 2. However, this model should be used with low confidence because it was originally designed for ferritic-pearlitic steels, and it is reported that the error from this model can be up to 34 K[77]. Thus, the ITT criterion for this design to select composition with good ductility at low temperature is set to be 0 °C to avoid over-filtering.

The chemical composition determines the weldability by influencing the hardenability and phase transformations during the welding process. Carbon plays a crucial role in weldability and has two major effects. Firstly, high carbon content leads to carbide precipitation during the AM process and increases the freezing range (the difference between the liquidus and solidus temperatures), which may initiate cracking through hot tearing effects[78]. Secondly, it causes an increase in hardenability and thus lowers the ductility[79]. The low carbon content of HSLA steel makes it a suitable candidate material for additive manufacturing. In this study, the ability to avoid hot and cold cracking for different compositions is evaluated by calculating the freezing range and the location in the Graville diagram[23], as shown in Fig. 2.

Hot cracking occurs near the solidus temperature where the liquid exists. A reduced freezing range is desirable to avoid hot cracking during additive manufacturing[80,81]. In this study, the freezing range is $T_{80\%liquid} - T_{20\%liquid}$ (the difference between temperatures with 80% and 20% liquid), and the equilibrium freezing range is calculated based on the TCFE9 database of the Thermo-Calc software. The allowable maximum freezing range for compositions with good weldability is set to be 13 K.

Cold cracking occurs when the weld has cooled down to room temperature, which is also called hydrogen-induced cracking (HIC). As a phenomenological method, the Graville diagram is very useful in determining the ability to avoid HIC[23]. If the alloy composition locates in Zone I of the Graville diagram, cold cracking only occurs when the hydrogen content is very high, and weldability is good. In contrast, compositions in Zone II or Zone III have a medium or high susceptibility to HIC, respectively[82], and the details of the Zone in Graville diagram and the location of different steels are illustrated by Caron, J[82]. An alloy with good weldability should satisfy the following equation to avoid cold cracking:

$$0 \leq -0.0515 \cdot CE + 0.127 - C \qquad (17)$$

where C is the carbon content of steel in weight percent, CE = C + (Mn+Si)/6 + (Ni+Cu)/15 + (Cr+Mo+V)/5 is the carbon equivalent (CE) of the steel in wt.%.

**Screening strategy, analysis, and verification**

The initial composition range and the screening range listed in Table 1 are employed for high-throughput calculations. The screening range spans a broader composition space in comparison with the initial composition range provided by the vendor. The screening composition range was determined to ensure that in the screening range, percent of compositions meeting all property requirements exhibit a peak or plateau for each composition screening range so that we do not miss the possible optimized composition space. Since there are nine elements whose composition needs to be optimized, it implies that there are nine variables with a certain range that needs to be considered in the mathematical space for sampling. The sampling space will have an exponential increase associated with a broad composition range for each element and thus require a huge sampling size to ensure that the analysis is based on a sufficient number of calculations. For example, if we discover the optimized composition for all components in the screening range that we defined in one time, it is found that such a multi-dimensional composition space is $1.7 \times 10^6$ times larger than the initial composition space, and $1.7 \times 10^6$ is the product of the ratios listed in Table 1. As a result, it is challenging to screen a sufficient number of compositions to represent the whole screening space. As a mitigation method to reduce the computational load, we optimized the composition for each element one by one. Take carbon as an example, we sampled 50,000 compositions from the screening range of carbon and initial composition range for the rest of the elements using the Latin hypercube sampling (LHS) approach[83] with a uniform distribution. The same procedure was repeated for all elements, and finally, 450, 000 compositions were sampled. This method requires much fewer calculations during the screening process, while still effectively cover the required composition space for discovering the optimized composition. The yield strength, ITT, and weldability of these samples were calculated with the aforementioned ICME framework to identify the influence of each element on the microstructure-property relationship. During the optimization process, we utilized nine cores and finished the simulation in less than two days, which proves that the efficiency of this computational framework is high. Further, the composition was optimized such that it maximized the possibility of a successful build, which could satisfy all the requirements for yield strength, weldability, and low-temperature ductility after post-heat treatment.

Once the optimized composition was determined, 50,000 compositions within the uncertainty range of the initial nominal composition and the optimized nominal composition were sampled using the Latin hypercube sampling approach[83] following a uniform distribution, respectively. Later, the probability analysis on successful additive manufacturing was performed, and the improvement in the optimized composition compared with the initial composition was evaluated.

Moreover, building surrogate models to overcome the computational challenge in optimization has been proved successful and efficient[84,85]. As a result, the machine learning approach was used to build surrogate models to optimize the alloy composition[86,87], all models were evaluated using 10-fold cross-validation[88], and the statistical analyses, such as Spearman's rank correlation and Sobol's indices, were performed to understand the influence of elements on the properties[89–93]. An optimal composition set with 69 compositions was found, and some compositions are close to the composition determined in this work. The details are given in the supplementary note.

## Data availability

The data that support the findings of this study are available from the corresponding author upon reasonable request.

## Code availability

The code that supports the findings of this study is available from the corresponding author upon reasonable request.

## Acknowledgments

The authors are grateful for helpful discussions with Dr. Soumya Sridar, Mr. Rafael Tomás Rodríguez De Vecchis, and Mr. Noah Sargent. The financial support received from the Office of Naval Research, Office of Naval Research (ONR) Additive Manufacturing Alloys for Naval Environments (AMANE) program (Contract No.: N00014-17-1-2586) is gratefully acknowledged for performing the current research.

# Figures

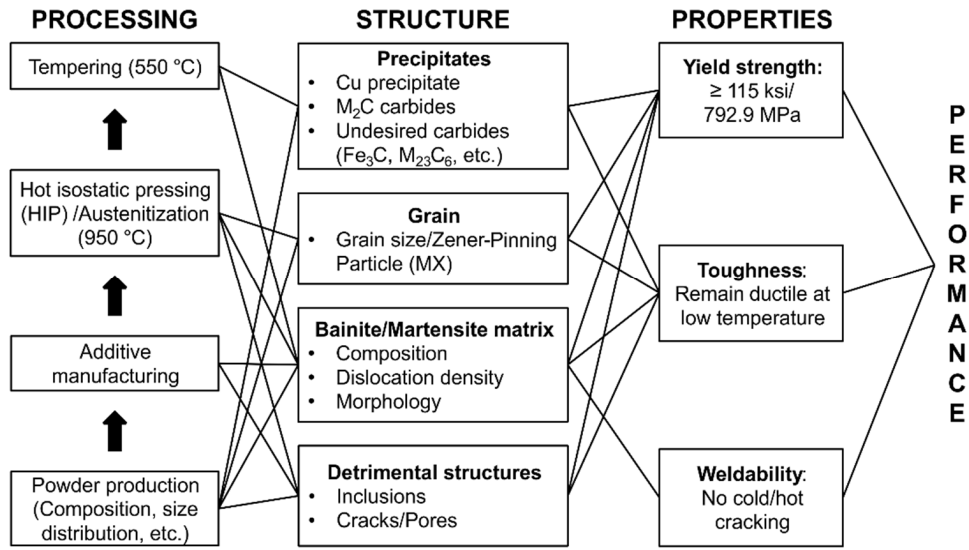

**Fig. 1 Systems design chart for AM HSLA-115.**

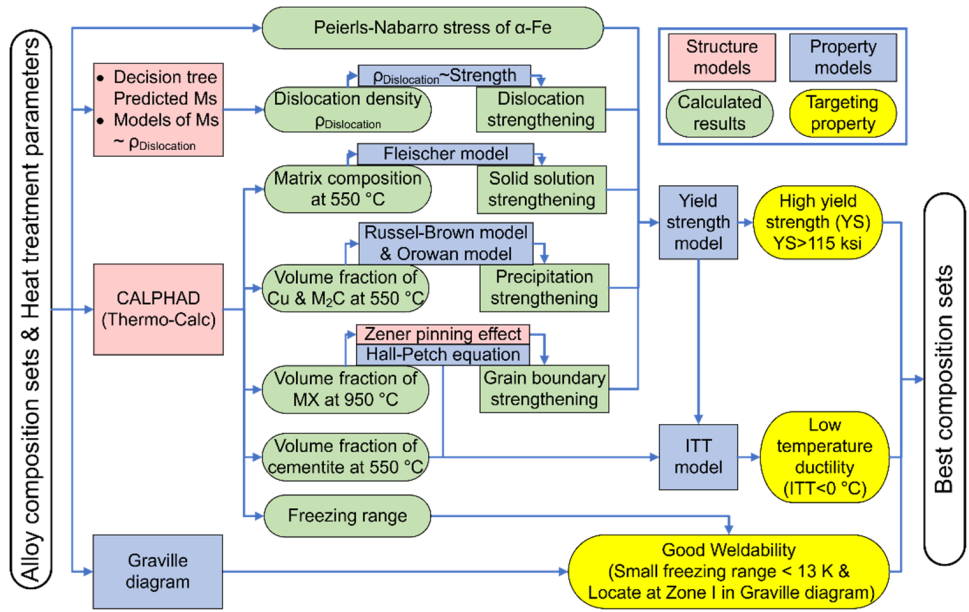

**Fig. 2 ICME modeling workflow for HSLA-115 steel composition design.** The pink box denotes structure models predicting features such as phase fraction of different phases, dislocation density based on composition and heat treatment process; the blue box denotes the property models which can simulate the strength, freezing range, etc. based on structure and compositions; the green box denotes the calculated property or structural information from the models; the yellow box denotes the target properties.

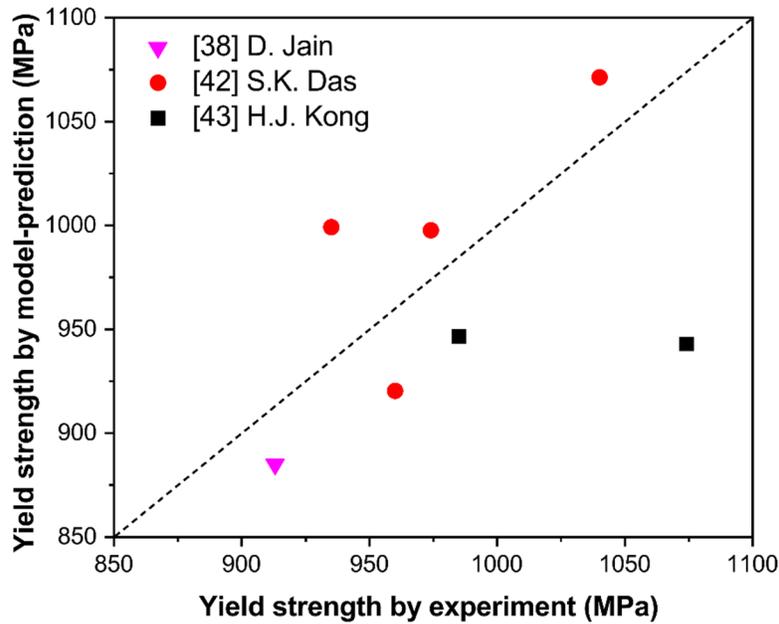

**Fig. 3 Comparison of the yield strength by model prediction and the experiments.** The model-predicted value is equal to the experimental value if the symbol is located on the dashed diagonal line.

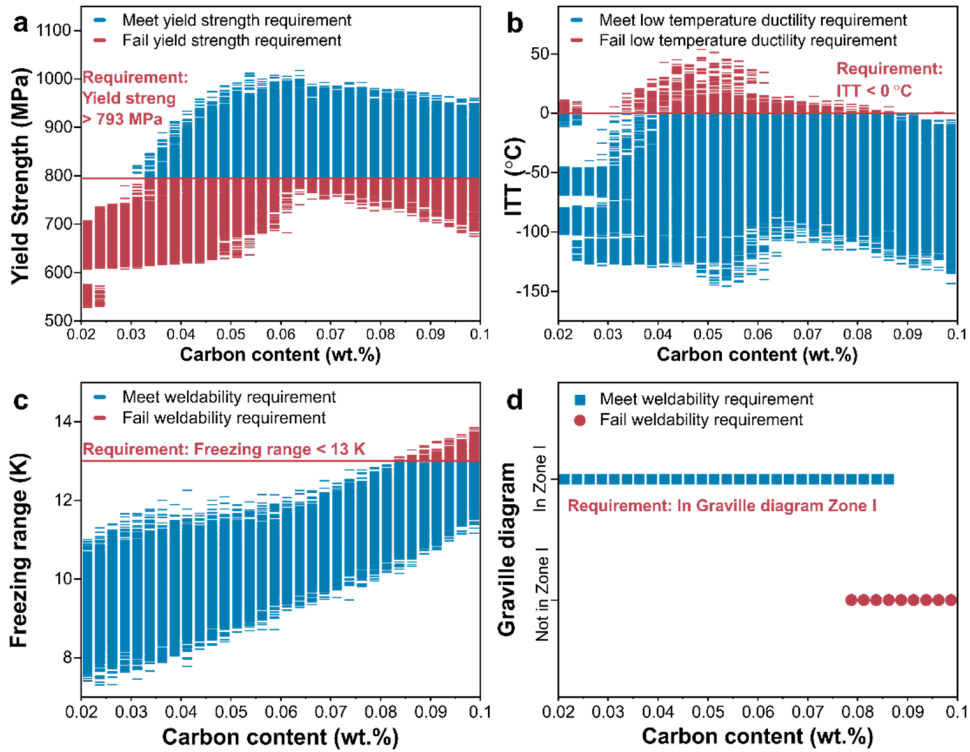

**Fig. 4 Variation in properties due to the change in carbon content.** Trend analysis on (a) yield strength, (b) ITT, (c) freezing range, and (d) Graville diagram location. The compositions meeting with the requirements are in blue. The compositions that failed the property requirements are in red.

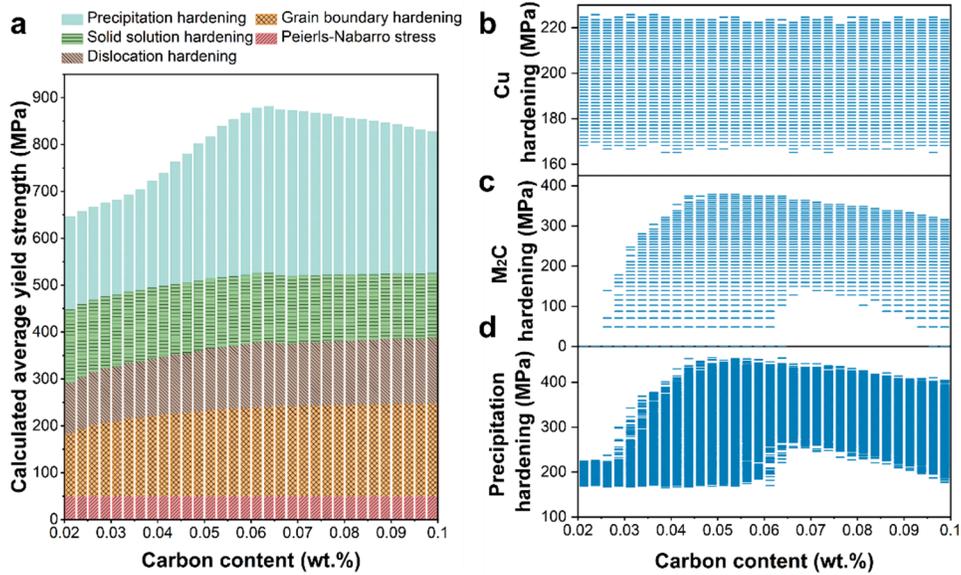

**Fig. 5 Predicted yield strength and the contribution from different strengthening mechanisms versus carbon content.** (a) Average strength from different strengthening effect versus carbon content, precipitation strengthening effect from (b) Cu precipitation, (c) M2C precipitation, and (d) sum of Cu and M2C precipitates.

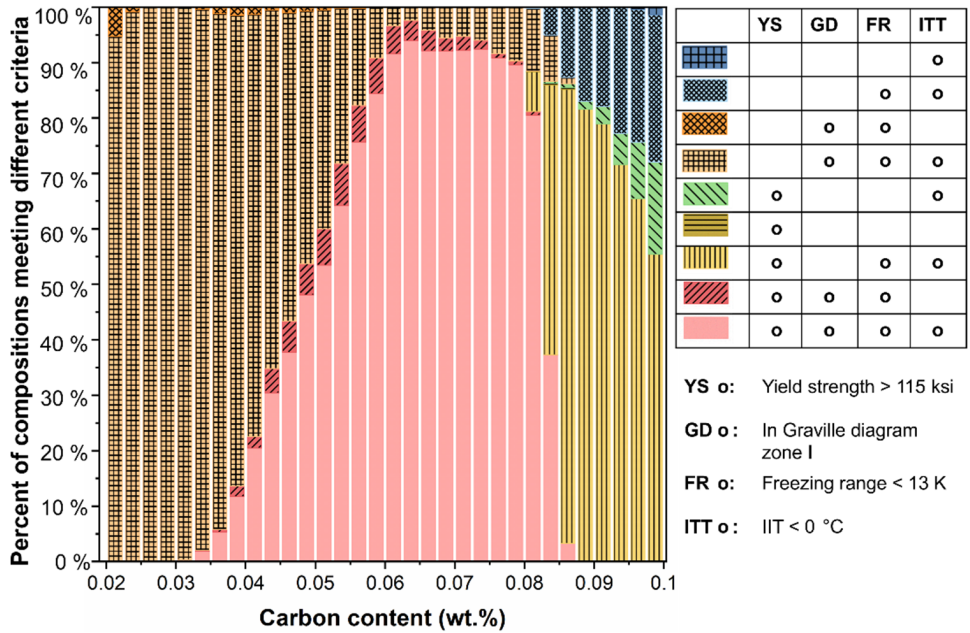

**Fig. 6 Optimization of carbon content by visualizing the percentages of compositions meeting different criteria.** Pink color without pattern filling: The percentage of compositions with yield strength higher than 115 ksi, good weldability, and ITT lower than 0 K. The meaning of other color and pattern-filled bars can be understood in a similar way based on the table in the figure, and groups smaller than 0.1 % are not listed for better illustration.

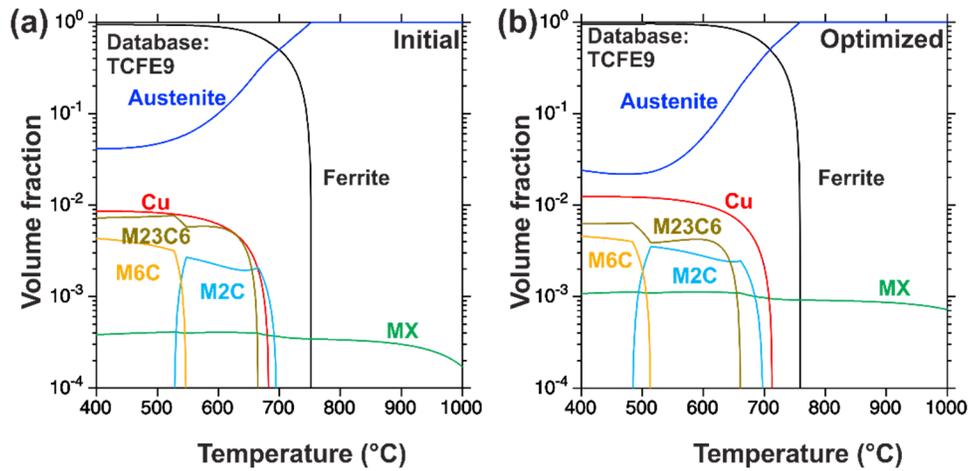

**Fig. 7 Equilibrium phase fraction plots as a function of temperature.** Diagrams of (a) initial and (b) optimized compositions calculated using the TCFE9 database.

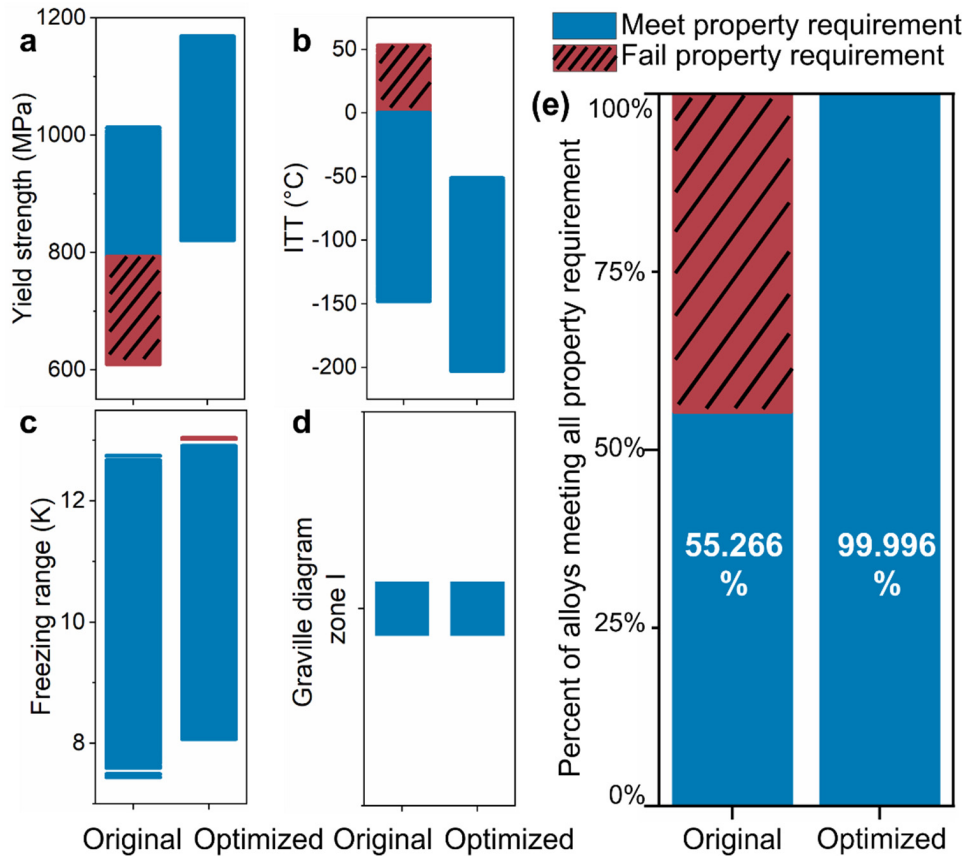

**Fig. 8 Distribution of calculated key properties of initial and optimized composition within their uncertainty range.** (a) yield strength, (b) ITT, (c) freezing range, and (d) location at Graville diagram. (e) Percentage of alloys meeting the criteria of initial composition and optimized composition. The ones meeting with the requirement are in blue without stripe pattern. The ones failed to match with the requirement are in red with stripe pattern.

Page **22** of **37**

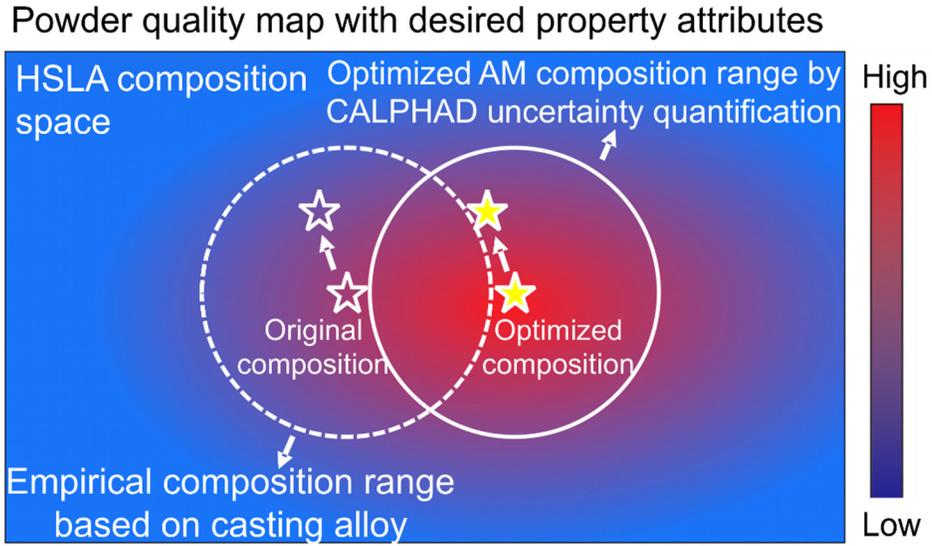

**Fig. 9 Conceptual graphic illustrating the improvement of composition with uncertainty after ICME optimization.** The color bar indicates the satisfactory of the powder composition, i.e., powder quality, with the potential to match the requirements of the design target.

# Tables

Table 1 The initial composition range (wt.%), screening range (wt.%), and their ratio for different elements in the manufactured AM powder for HSLA-115 steel

|  | Fe | C | Cr | Cu | Mn | Nb | Mo | Ni | Si | Al |
|---|---|---|---|---|---|---|---|---|---|---|
| Initial composition range | Bal. | 0.053 ± 0.025 | 0.66 ± 0.10 | 1.27 ± 0.15 | 0.98 ± 0.20 | 0.03 ± 0.01 | 0.57 ± 0.10 | 3.43 ± 0.20 | 0.225 ± 0.125 | 0.03 ± 0.01 |
| Screening composition range | Bal. | 0.06 ± 0.04 | 0.6 ± 0.5 | 1.25 ± 0.45 | 1.15 ± 0.95 | 0.055 ± 0.045 | 0.7 ± 0.5 | 3.5 ± 1.5 | 0.25 ± 0.25 | 0.055 ± 0.045 |
| Ratio (Screening range/ Initial composition range) |  | 1.6 | 5 | 3 | 4.75 | 4.5 | 5 | 7.5 | 2 | 4.5 |

Table 2 Summary of the influence of elements in HSLA-115 on the key properties[1,2], for each screening range

|  | C | Cr | Cu | Mn | Nb | Mo | Ni | Si | Al |
|---|---|---|---|---|---|---|---|---|---|
| Resistance to cold cracking | ↓ | ↓ | O | ↓ | O | ↓ | ↓ | O | O |
| Resistance to hot cracking | ↓ | O | O | ↓ | ↓ | ↓ | ↑ | ↓ | ↓ |
| Ductility at low temperature | ↓↑ | ↓↑ | ↓ | ↑ | ↑ | ↓↑ | O | ↑ | O |
| Yield strength | ↑↓ | ↑↓ | ↑ | ↓ | ↑ | ↑↓ | O | ↓ | O |
| Cu hardening | O | O | ↑ | ↓ | O | O | ↓ | ↑ | O |
| M$_2$C hardening | ↑↓ | ↑↓ | O | O | ↓ | ↑↓ | O | ↓ | O |
| Solid solution hardening | ↓ | ↑ | O | ↓ | ↑ | ↑ | ↑ | O | O |
| Dislocation hardening | ↑ | ↑ | O | ↑ | O | ↑ | ↑ | O | O |
| Grain boundary hardening | ↑ | O | O | O | ↑ | O | O | O | O |

1: Notations in the table are:
  ↑: The increase in the component is beneficial to the property;
  ↓: The increase in the component is detrimental to the property;
  O: The increase in the component has no obvious effect on the property;
  ↑↓: The increase in the component is beneficial to the property first, and then detrimental to the property;
  ↓↑: The increase in the component is detrimental to the property first, and then beneficial to the property.
2: Examples of analysis has been illustrated in Fig. 4 and Fig. 5 for carbon

Table 3 Comparison of initial composition and optimized composition (wt.%)

| Element | Fe | C | Cr | Cu | Mn | Nb | Mo | Ni | Si | Al |
|---|---|---|---|---|---|---|---|---|---|---|
| Initial composition | Bal. | 0.053 ± 0.025 | 0.66 ± 0.1 | 1.27 ± 0.15 | 0.98 ± 0.2 | 0.03 ± 0.01 | 0.57 ± 0.1 | 3.43 ± 0.2 | 0.225 ± 0.125 | 0.03 ± 0.01 |
| Optimized composition | Bal. | 0.057 ± 0.025 | 0.525 ± 0.1 | 1.55 ± 0.15 | 0.5 ± 0.2 | 0.08 ± 0.01 | 0.57 ± 0.1 | 3.43 ± 0.2 | 0.125 ± 0.125 | 0.03 ± 0.01 |

Table 4 Comparison of model-predicted key properties of the initial and optimized nominal compositions

| Calculated properties | Yield strength | ITT | Freezing range | Graville diagram |
|---|---|---|---|---|
| Initial composition | 873 MPa | -15°C | 10.10 K | Zone I |
| Optimized composition | 1076 MPa | -100°C | 10.24 K | Zone I |

## Supplementary Figures:

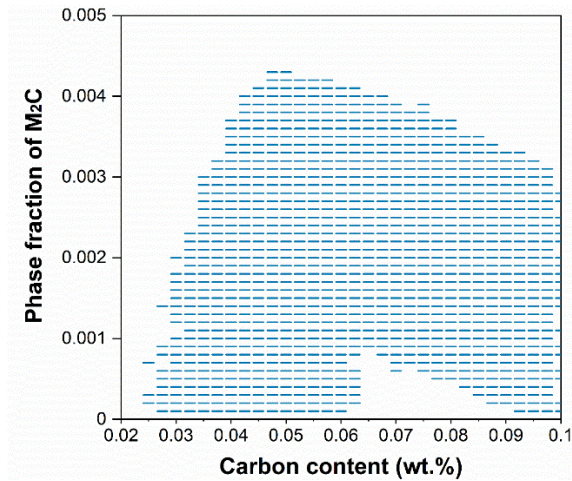

Supplementary Figure 1: Change of $M_2C$ fraction with different carbon content within the screening range and different other elements within their initial composition range

## Supplementary notes:

The surrogate models are computationally cheap and can be used to find the optimal solution[84,85]. Thus, we applied the surrogate model to find the optimized compositions following the steps listed in Supplementary Table 1, where are details are given below.

Supplementary Table 1 Steps of determining the optimal composition using surrogate models

| Step | Description |
| --- | --- |
| 1 | Sample 500, 000 samples using LHS method, and calculate the properties of interest using CALPHAD-based ICME modeling framework |
| 2 | Build surrogate models using gradient boosting algorithms and 500, 000 compositions calculated in step 1, evaluate models using 10-fold cross-validation |
| 3 | Perform Spearman's rank correlation analysis and calculate Sobol's indices to reveal the elements influence and the importance of element in the uncertainty of critical properties, and determine the screening compositions |
| 4 | Prepare 100 000 compositions using the LHS method within the uncertainty range of each screened compositions and calculate the property of interests & Find the optimal compositions that all 10, 000 compositions within its uncertainty range meet all property requirements. |

We sampled 500,000 samples from the screening composition range listed in Supplementary Table 2 using the LHS method. The properties of interest of all compositions have been calculated using the CALPHAD-based modeling framework (Fig. 2 in manuscript).

Supplementary Table 2 Summary composition range in wt.% used for building the surrogate model

| Fe | C | Cr | Cu | Mn | Nb | Mo | Ni | Si | Al |
|---|---|---|---|---|---|---|---|---|---|
|  | 0.02 | 0.1 | 0.8 | 0.2 | 0.01 | 0.2 | 2 | 0 | 0.01 |
| Bal. | - | - | - | - | - | - | - | - | - |
|  | 0.1 | 1.1 | 1.7 | 2.1 | 0.1 | 1.2 | 5 | 0.5 | 0.1 |

The criteria for the Graville diagram is a linear regression, so there is no need to build a surrogate model for it. Four surrogate models predicting the yield strength, freezing range, impact transition temperature (ITT), and the phase fraction of cementite were built by using the 500,000 calculated results and gradient boosting (GB) algorithm[86,87]. All models were evaluated using 10-fold cross-validation[88]. During the cross-validation, the 500,000 samples were randomly split into ten subsets, and the model is fitted with nine subgroups (training dataset) and tested with the remaining subgroup (testing dataset). After training and testing ten times, the average performance metrics such as the mean absolute error (MAE), root mean square error (RMSE), and coefficient of determination ($R^2$) were calculated using the following Eqs. (1-3), respectively:

$$\text{MAE} = \frac{\sum_{i=1}^{n}|y_i^P - y_i^S|}{n} \quad (1)$$

$$\text{RMSE} = \sqrt{\frac{\sum_{i=1}^{n}(y_i^P - y_i^S)^2}{n}} \quad (2)$$

$$R^2 = 1 - \frac{\sum_{i=1}^{n}(y_i^P - y_i^S)^2}{\sum_{i=1}^{n}(y_i^P - \overline{y^P})^2} \quad (3)$$

where $n$ is the number of data points in the testing dataset, $y_i^P$ and $y_i^S$ are CALPHAD-based model framework predicted property and surrogate model predicted results of the datapoint $i$, respectively. $\overline{y^P} = \frac{\sum_{i=1}^{n} y_i^P}{n}$ is the mean value of $y_i^P$ in the testing dataset.

Supplementary Table 3 lists the performance of the surrogate models. The MAE represents the average of absolute error of the model prediction, while the RMSE is the standard deviation of the error introduced by the surrogate model that penalizes the samples with large error. $R^2$ measures to what extent the variance of a dependent variable can be explained by the variables in a regression model. For yield strength, ITT, and freezing range, more than 90 % variance of results can be explained by the model, and their MAE and RMSE are acceptable. But the model prediction of the cementite phase fraction only has low MAE and RMSE, which means a small error made by the model. But the $R^2$ is close to 0, indicating the variance cannot be well-explained.

Supplementary Table 3 Average of the MAE, RMSE, and $R^2$ of the 10-fold cross-validation

| Properties of interest | MAE | RMSE | $R^2$ |
|---|---|---|---|
| Yield strength (MPa) | 33.80 | 44.52 | 0.93 |
| ITT (°C) | 16.00 | 20.76 | 0.94 |
| Freezing range (K) | 0.37 | 0.49 | 0.97 |
| Cementite fraction | 0.0000 | 0.0002 | 0.00 |

In order to reveal the element influence on the properties of interest, we first performed the Spearman's rank correlation[89] of the properties predicted by CALPHAD-based modeling framework and the composition based on 500,000 calculated results. The Spearman's correlation coefficient $\rho$ and its $p$ are listed in Supplementary Table 4. The coefficient $\rho$ measures the strength of the monotonic relationship, and its value ranges from -1 to 1. A positive/negative value represents a positive/negative monotonic relationship, while 0 indicates no monotonic relationship between two variables. If the absolute value of $\rho$ is smaller than 0.4, the monotonic relationship is weak, if $\rho$ is higher than 0.6, the relationship is strong, and there is a moderate monotonic relationship if $\rho$ is located between 0.4 to 0.6[90]. The associated $p$ indicates the chance of the null hypothesis is true, and the null hypothesis in calculating $\rho$ is that we can get the same $\rho$ when the two variables are not correlated. If the $p$ is smaller than 0.05, one can conclude that calculated Spearman's coefficient is statistically significant[91].

Supplementary Table 4 Spearman's correlation coefficient $\rho$ and its *p*-value of the composition and properties of interest

|  | Graville diagram | | Freezing range | | Yield strength | | ITT | | Cementite fraction | |
|---|---|---|---|---|---|---|---|---|---|---|
|  | $\rho$ | *p*-value | $\rho$ | *p*-value | $\rho$ | *p*-value | $\rho$ | *p*-value | $\rho$ | *p*-value |
| C  | -0.71 | 0.00 | 0.58  | 0.00 | 0.31  | 0.00 | 0.09  | 0.00 | 0.06  | 0.00 |
| Cr | -0.08 | 0.00 | -0.04 | 0.00 | -0.31 | 0.00 | -0.32 | 0.00 | -0.08 | 0.00 |
| Cu | -0.03 | 0.00 | 0.01  | 0.00 | 0.21  | 0.00 | 0.15  | 0.00 | 0.00  | 0.74 |
| Mn | -0.13 | 0.00 | 0.18  | 0.00 | -0.06 | 0.00 | -0.07 | 0.00 | 0.03  | 0.00 |
| Nb | 0.00  | 0.63 | 0.16  | 0.00 | 0.48  | 0.00 | -0.69 | 0.00 | -0.01 | 0.00 |
| Mo | -0.08 | 0.00 | 0.35  | 0.00 | 0.30  | 0.00 | 0.22  | 0.00 | -0.07 | 0.00 |
| Ni | -0.08 | 0.00 | -0.50 | 0.00 | -0.05 | 0.00 | -0.09 | 0.00 | -0.03 | 0.00 |
| Si | -0.04 | 0.00 | 0.24  | 0.00 | -0.01 | 0.00 | -0.02 | 0.00 | -0.01 | 0.00 |
| Al | 0.00  | 0.39 | 0.06  | 0.00 | 0.00  | 0.76 | 0.00  | 0.03 | 0.00  | 0.25 |

Take carbon as an example. The carbon content is correlated to all properties since they all have a non-zero $\rho$, and *p*-values are all 0. If the carbon is increasing, there is a robust monotonic trend to shift the Graville diagram from Zone I (1) to Zone II/III (0). The freezing range and yield strength have a moderate monotonic trend to increase when carbon is added to the alloy. Moreover, ITT and cementite fraction only show a weak positive monotonic relationship with carbon content. The influence of other elements can also be analyzed similarly. It is noted that Al, have no monotonic relationship with Graville diagram, yield strength, ITT, and Cementite, and only has a weak positive relationship with a freezing range.

In order to reduce the dimension of screening composition space, the global sensitivity analysis was performed to reveal the contribution of each element on the variance of the property of interest by calculating the Sobol's indices[92,93]. The first-order indices (S1) and total indices (ST) measures the contribution to output variance by a single input alone and the total contributions of an input that include interactions with other variables. The S1, ST, and their 95 % confidence intervals (Conf) of each input in the surrogate model have been calculated and listed in Supplementary Table 5. The Sobol's indices of Al are close to 0 in all surrogate models. Moreover, Al is not included in the equation for determining the location in the Graville diagram. As a result, during the composition screening process, Al is fixed as the original composition to reduce the number of compositions during the optimization. Supplementary Table 6 shows the screening grids and the uncertainty ranges, and 5,334,336 nominal compositions will be screened.

Supplementary Table 5 Sensitivity analysis with first-order and total Sobol's indices and for each element used in the surrogate models [1,2,3]

|    | Freezing range | | Yield strength | | ITT | |
|----|---------------|---------------|---------------|---------------|---------------|---------------|
|    | S1±Conf | ST±Conf | S1±Conf | ST±Conf | S1±Conf | ST±Conf |
| C  | 0.341±0.017 | 0.412±0.013 | 0.146±0.014 | 0.382±0.016 | 0.012±0.011 | 0.176±0.008 |
| Cr | **0.002±0.001** | **0.002±0.000** | 0.156±0.014 | 0.335±0.013 | 0.136±0.011 | 0.256±0.009 |
| Cu | **0.000±0.002** | **0.003±0.000** | 0.037±0.006 | 0.039±0.001 | 0.023±0.004 | 0.025±0.001 |
| Mn | 0.028±0.006 | 0.043±0.002 | **0.002±0.002** | **0.006±0.000** | **0.007±0.002** | **0.009±0.000** |
| Nb | 0.025±0.004 | 0.025±0.001 | 0.219±0.013 | 0.227±0.006 | 0.526±0.020 | 0.535±0.014 |
| Mo | 0.137±0.011 | 0.166±0.006 | 0.138±0.018 | 0.369±0.017 | 0.074±0.012 | 0.231±0.011 |
| Ni | 0.304±0.014 | 0.386±0.011 | **0.002±0.002** | **0.004±0.000** | **0.008±0.003** | 0.011±0.000 |
| Si | 0.062±0.007 | 0.063±0.002 | **0.001±0.001** | **0.001±0.000** | **0.000±0.001** | **0.001±0.000** |
| Al | **0.005±0.002** | **0.006±0.000** | **0.000±0.000** | **0.000±0.000** | **0.000±0.000** | **0.000±0.000** |

1 The Sobol's indices for cementite phase fraction are not calculated, due to invalid value error encountered in the calculation (Standard deviation in cementite fraction is 0 during calculation).
2 In Graville diagram, *Zone I is when* $0 \leq -0.0515 \cdot CE + 0.127 - C$, $CE = C + (Mn+Si)/6 + (Ni+Cu)/15 + (Cr+Mo+V)/5$. Only Al and Nb are not influencing the results.
3 Indices smaller than 0.01 are marked in bold font

Supplementary Table 6 Screening compositions and the uncertainty range

|  | Fe | C | Cr | Cu | Mn | Nb | Mo | Ni | Si | Al |
|---|---|---|---|---|---|---|---|---|---|---|
| Screening composition | Bal. | 0.045, 0.050, …, 0.075 | 0.2, 0.3, …, 1.0 | 1.0, 1.15, …, 1.45 | 0.4, 0.6, …, 1.8 | 0.02, 0.03, …, 0.08 | 0.3, 0.4, …, 1.1 | 2.2, 2.4, …, 4.8 | 0.125, 0.250, 0.375 | 0.03 |
| Uncertainty range | Bal. | ± 0.025 | ± 0.1 | ± 0.15 | ± 0.2 | ± 0.01 | ± 0.1 | ± 0.2 | ± 0.125 | ± 0.01 |

For each grid defined by Supplementary Table 6, 5,000 compositions have been generated using the LHS approach within the composition uncertainty range and calculated using the surrogate models. For example, for one composition defined by the smallest value in all elements, 5,000 compositions were obtained from C:0.045 ± 0.025, Cr:0.2 ± 0.1, Cu: 1.0± 0.15, Mn:0.4± 0.2, Nb: 0.02± 0.01, Mo:0.3± 0.1, Ni:2.2± 0.2, Si:0.125± 0.125, Al:0.03± 0.01, and their properties were calculated. The mean and standard deviation of freezing range, yield strength an ITT, and the percent of samples meeting the all property criteria (criteria are listed in the manuscript) of 5,000 compositions were also calculated. This procedure has been performed for all combinations listed in Table S2-6. The compositions that have a 100 % chance to meet all property requirements have been stored and used for another screening. In the second screening, we generated 100,000 compositions rather than 5,000 compositions and modified the property criteria with the surrogate model RMSE to increase the reliability. The criteria are Graville diagram located in zone I, freezing range < 13 – 0.5 K, yield strength > 793 + 45 MPa, ITT < 0 – 21 K, and cementite fraction < 0.0001 (Cementite is bad for low temperature ductility). Finally, 69 compositions that enable all 100,000 compositions with their uncertainty range to meet all property requirements have been found (Supplementary Table 7). It is clear that lots of the compositions listed in Supplementary Table 7 are close to what we have designed in the manuscript.

Supplementary Table 7 Optimal compositions determined using surrogate models *

| C | Cr | Cu | Mn | Nb | Mo | Ni | Si | Al | Yield strength (MPa) | | Freezing range (K) | | ITT (°C) | |
|---|---|---|---|---|---|---|---|---|---|---|---|---|---|---|
| | | | | | | | | | Mean | STD | Mean | STD | Mean | STD |
| 0.055 | 0.5 | 1.45 | 0.4 | 0.08 | 0.5 | 4 | 0.125 | 0.03 | 984 | 45 | 9.1 | 1.1 | -137 | 20 |
| 0.055 | 0.5 | 1.45 | 0.4 | 0.08 | 0.5 | 4.6 | 0.125 | 0.03 | 982 | 46 | 9.1 | 1.2 | -141 | 20 |
| 0.055 | 0.3 | 1.45 | 0.4 | 0.08 | 0.6 | 4.2 | 0.125 | 0.03 | 1035 | 51 | 9.4 | 1.1 | -113 | 22 |
| 0.055 | 0.3 | 1.45 | 0.4 | 0.08 | 0.6 | 4.4 | 0.125 | 0.03 | 1035 | 52 | 9.4 | 1.2 | -115 | 22 |
| 0.055 | 0.4 | 1.45 | 0.4 | 0.07 | 0.6 | 3.8 | 0.125 | 0.03 | 1029 | 53 | 9.6 | 1.0 | -94 | 20 |
| 0.055 | 0.4 | 1.45 | 0.4 | 0.07 | 0.6 | 4 | 0.125 | 0.03 | 1030 | 53 | 9.3 | 1.1 | -95 | 20 |
| 0.055 | 0.4 | 1.45 | 0.4 | 0.07 | 0.6 | 4.2 | 0.125 | 0.03 | 1030 | 53 | 9.2 | 1.1 | -96 | 21 |
| 0.055 | 0.4 | 1.3 | 0.4 | 0.07 | 0.6 | 4 | 0.125 | 0.03 | 1016 | 54 | 9.4 | 1.1 | -100 | 21 |
| 0.055 | 0.4 | 1.45 | 0.6 | 0.07 | 0.6 | 4 | 0.125 | 0.03 | 1027 | 54 | 9.5 | 1.1 | -97 | 21 |
| 0.055 | 0.4 | 1.45 | 0.4 | 0.07 | 0.6 | 4.4 | 0.125 | 0.03 | 1030 | 54 | 9.1 | 1.2 | -97 | 21 |
| 0.055 | 0.4 | 1.45 | 0.6 | 0.07 | 0.6 | 4.2 | 0.125 | 0.03 | 1027 | 54 | 9.4 | 1.1 | -98 | 21 |
| 0.055 | 0.4 | 1.3 | 0.4 | 0.07 | 0.6 | 4.2 | 0.125 | 0.03 | 1016 | 54 | 9.2 | 1.1 | -101 | 21 |
| 0.055 | 0.4 | 1.45 | 0.4 | 0.08 | 0.6 | 4 | 0.125 | 0.03 | 1046 | 54 | 9.5 | 1.1 | -109 | 21 |
| 0.055 | 0.4 | 1.45 | 0.4 | 0.08 | 0.6 | 4.2 | 0.125 | 0.03 | 1046 | 54 | 9.4 | 1.1 | -110 | 21 |
| 0.055 | 0.4 | 1.45 | 0.4 | 0.07 | 0.6 | 4.6 | 0.125 | 0.03 | 1029 | 54 | 9.2 | 1.2 | -98 | 21 |
| 0.055 | 0.4 | 1.45 | 0.6 | 0.07 | 0.6 | 4.4 | 0.125 | 0.03 | 1027 | 54 | 9.4 | 1.2 | -99 | 21 |
| 0.055 | 0.4 | 1.3 | 0.4 | 0.07 | 0.6 | 4.4 | 0.125 | 0.03 | 1016 | 54 | 9.1 | 1.2 | -102 | 21 |
| 0.055 | 0.4 | 1.45 | 0.6 | 0.08 | 0.6 | 4 | 0.125 | 0.03 | 1043 | 54 | 9.7 | 1.1 | -111 | 21 |
| 0.055 | 0.4 | 1.45 | 0.4 | 0.08 | 0.6 | 4.4 | 0.125 | 0.03 | 1045 | 54 | 9.3 | 1.2 | -112 | 21 |
| 0.055 | 0.4 | 1.3 | 0.4 | 0.08 | 0.6 | 4 | 0.125 | 0.03 | 1033 | 55 | 9.5 | 1.1 | -113 | 21 |
| 0.055 | 0.4 | 1.3 | 0.6 | 0.07 | 0.6 | 4.4 | 0.125 | 0.03 | 1012 | 55 | 9.4 | 1.2 | -104 | 22 |
| 0.055 | 0.4 | 1.3 | 0.4 | 0.07 | 0.6 | 4.6 | 0.125 | 0.03 | 1015 | 55 | 9.1 | 1.2 | -103 | 22 |
| 0.055 | 0.4 | 1.3 | 0.4 | 0.08 | 0.6 | 4.2 | 0.125 | 0.03 | 1033 | 55 | 9.4 | 1.1 | -115 | 22 |
| 0.055 | 0.4 | 1.3 | 0.4 | 0.07 | 0.6 | 4.8 | 0.125 | 0.03 | 1015 | 55 | 9.2 | 1.2 | -104 | 22 |
| 0.055 | 0.4 | 1.3 | 0.4 | 0.08 | 0.6 | 4.4 | 0.125 | 0.03 | 1033 | 55 | 9.3 | 1.2 | -116 | 22 |
| 0.055 | 0.4 | 1.3 | 0.6 | 0.08 | 0.6 | 4.2 | 0.125 | 0.03 | 1030 | 56 | 9.6 | 1.1 | -117 | 22 |
| 0.055 | 0.4 | 1.3 | 0.4 | 0.08 | 0.6 | 4.6 | 0.125 | 0.03 | 1032 | 56 | 9.3 | 1.2 | -117 | 22 |
| 0.055 | 0.4 | 1.3 | 0.6 | 0.08 | 0.6 | 4.4 | 0.125 | 0.03 | 1029 | 56 | 9.6 | 1.2 | -118 | 22 |
| 0.055 | 0.4 | 1.15 | 0.4 | 0.08 | 0.6 | 4.2 | 0.125 | 0.03 | 1022 | 57 | 9.4 | 1.1 | -120 | 22 |
| 0.055 | 0.4 | 1.15 | 0.4 | 0.08 | 0.6 | 4.4 | 0.125 | 0.03 | 1021 | 57 | 9.3 | 1.2 | -122 | 23 |
| 0.055 | 0.4 | 1.15 | 0.6 | 0.08 | 0.6 | 4.2 | 0.125 | 0.03 | 1019 | 57 | 9.6 | 1.1 | -122 | 23 |
| 0.055 | 0.4 | 1.15 | 0.6 | 0.08 | 0.6 | 4.4 | 0.125 | 0.03 | 1017 | 58 | 9.5 | 1.2 | -124 | 23 |
| 0.055 | 0.5 | 1.45 | 0.4 | 0.08 | 0.6 | 4 | 0.125 | 0.03 | 1026 | 64 | 9.5 | 1.1 | -119 | 24 |
| 0.055 | 0.5 | 1.45 | 0.4 | 0.08 | 0.6 | 4.2 | 0.125 | 0.03 | 1027 | 65 | 9.3 | 1.1 | -120 | 24 |
| 0.055 | 0.5 | 1.45 | 0.4 | 0.08 | 0.6 | 4.4 | 0.125 | 0.03 | 1026 | 65 | 9.3 | 1.2 | -122 | 24 |
| 0.055 | 0.5 | 1.45 | 0.4 | 0.08 | 0.6 | 4.6 | 0.125 | 0.03 | 1025 | 65 | 9.3 | 1.2 | -122 | 24 |
| 0.055 | 0.3 | 1.45 | 0.4 | 0.08 | 0.7 | 4.4 | 0.125 | 0.03 | 1076 | 67 | 9.6 | 1.1 | -100 | 27 |
| 0.055 | 0.4 | 1.45 | 0.4 | 0.08 | 0.7 | 4.2 | 0.125 | 0.03 | 1084 | 77 | 9.7 | 1.1 | -97 | 28 |
| 0.055 | 0.4 | 1.3 | 0.4 | 0.08 | 0.7 | 4.2 | 0.125 | 0.03 | 1072 | 78 | 9.7 | 1.1 | -101 | 29 |
| 0.055 | 0.4 | 1.3 | 0.4 | 0.08 | 0.7 | 4.4 | 0.125 | 0.03 | 1071 | 78 | 9.5 | 1.1 | -102 | 30 |
| 0.05 | 0.4 | 1.45 | 0.4 | 0.08 | 0.5 | 4 | 0.25 | 0.03 | 999 | 39 | 9.4 | 1.1 | -128 | 17 |
| 0.05 | 0.4 | 1.45 | 0.4 | 0.08 | 0.5 | 4.2 | 0.25 | 0.03 | 999 | 39 | 9.3 | 1.1 | -130 | 18 |
| 0.05 | 0.4 | 1.45 | 0.4 | 0.08 | 0.5 | 3.6 | 0.125 | 0.03 | 998 | 40 | 9.6 | 0.9 | -125 | 17 |
| 0.05 | 0.4 | 1.45 | 0.4 | 0.08 | 0.5 | 4 | 0.125 | 0.03 | 998 | 40 | 8.8 | 1.1 | -128 | 17 |
| 0.05 | 0.4 | 1.45 | 0.6 | 0.08 | 0.5 | 4 | 0.25 | 0.03 | 997 | 40 | 9.6 | 1.1 | -131 | 17 |
| 0.05 | 0.4 | 1.45 | 0.4 | 0.08 | 0.5 | 4.4 | 0.25 | 0.03 | 998 | 40 | 9.3 | 1.2 | -131 | 18 |
| 0.05 | 0.4 | 1.45 | 0.4 | 0.08 | 0.5 | 4.2 | 0.125 | 0.03 | 998 | 40 | 8.7 | 1.1 | -129 | 18 |
| 0.05 | 0.4 | 1.45 | 0.6 | 0.08 | 0.5 | 3.8 | 0.125 | 0.03 | 995 | 40 | 9.2 | 1.0 | -129 | 18 |
| 0.05 | 0.4 | 1.45 | 0.4 | 0.08 | 0.5 | 4.4 | 0.125 | 0.03 | 997 | 40 | 8.7 | 1.2 | -131 | 18 |
| 0.05 | 0.4 | 1.45 | 0.6 | 0.08 | 0.5 | 4 | 0.125 | 0.03 | 995 | 40 | 9.0 | 1.0 | -130 | 18 |
| 0.05 | 0.4 | 1.3 | 0.4 | 0.08 | 0.5 | 3.6 | 0.125 | 0.03 | 985 | 40 | 9.7 | 0.9 | -131 | 18 |
| 0.05 | 0.4 | 1.45 | 0.6 | 0.08 | 0.5 | 4.2 | 0.125 | 0.03 | 995 | 41 | 9.0 | 1.1 | -132 | 18 |
| 0.05 | 0.4 | 1.45 | 0.6 | 0.08 | 0.5 | 4.4 | 0.125 | 0.03 | 994 | 41 | 9.0 | 1.2 | -133 | 18 |
| 0.05 | 0.4 | 1.45 | 0.8 | 0.08 | 0.5 | 4 | 0.125 | 0.03 | 991 | 41 | 9.2 | 1.0 | -132 | 18 |
| 0.05 | 0.2 | 1.45 | 0.4 | 0.08 | 0.6 | 4 | 0.125 | 0.03 | 1018 | 54 | 9.2 | 1.1 | -116 | 22 |
| 0.05 | 0.2 | 1.3 | 0.4 | 0.08 | 0.6 | 4.2 | 0.125 | 0.03 | 1004 | 54 | 9.1 | 1.1 | -122 | 22 |
| 0.05 | 0.2 | 1.45 | 0.8 | 0.08 | 0.6 | 4.2 | 0.125 | 0.03 | 1011 | 55 | 9.5 | 1.1 | -121 | 22 |
| 0.05 | 0.2 | 1.45 | 0.4 | 0.08 | 0.6 | 4.8 | 0.125 | 0.03 | 1016 | 55 | 9.1 | 1.2 | -121 | 23 |
| 0.05 | 0.3 | 1.45 | 0.4 | 0.08 | 0.6 | 3.8 | 0.125 | 0.03 | 1029 | 57 | 9.5 | 1.0 | -111 | 21 |
| 0.05 | 0.3 | 1.45 | 0.4 | 0.08 | 0.6 | 4 | 0.125 | 0.03 | 1029 | 57 | 9.2 | 1.1 | -112 | 21 |
| 0.05 | 0.3 | 1.45 | 0.4 | 0.08 | 0.6 | 4.2 | 0.125 | 0.03 | 1029 | 57 | 9.0 | 1.1 | -114 | 22 |
| 0.05 | 0.3 | 1.45 | 0.6 | 0.08 | 0.6 | 4 | 0.125 | 0.03 | 1027 | 57 | 9.4 | 1.0 | -114 | 22 |
| 0.05 | 0.3 | 1.45 | 0.4 | 0.08 | 0.6 | 4.4 | 0.125 | 0.03 | 1028 | 57 | 9.0 | 1.2 | -115 | 22 |
| 0.05 | 0.3 | 1.45 | 0.4 | 0.08 | 0.6 | 4.6 | 0.125 | 0.03 | 1027 | 58 | 9.0 | 1.2 | -116 | 22 |
| 0.05 | 0.3 | 1.45 | 0.6 | 0.08 | 0.6 | 4.4 | 0.125 | 0.03 | 1026 | 58 | 9.3 | 1.2 | -117 | 22 |
| 0.05 | 0.3 | 1.45 | 0.4 | 0.08 | 0.6 | 4.8 | 0.125 | 0.03 | 1027 | 58 | 9.1 | 1.2 | -117 | 23 |
| 0.05 | 0.4 | 1.45 | 0.4 | 0.08 | 0.6 | 3.8 | 0.125 | 0.03 | 1033 | 64 | 9.5 | 1.0 | -111 | 22 |
| 0.05 | 0.4 | 1.45 | 0.4 | 0.08 | 0.6 | 4.2 | 0.25 | 0.03 | 1033 | 64 | 9.5 | 1.1 | -115 | 23 |
| 0.05 | 0.4 | 1.45 | 0.4 | 0.08 | 0.6 | 4 | 0.125 | 0.03 | 1034 | 64 | 9.2 | 1.1 | -112 | 22 |